\documentclass[letterpaper]{ptephy}

\makeatletter
\newcommand{\figcaption}[1]{\def\@captype{figure}\caption{#1}}
\newcommand{\tblcaption}[1]{\def\@captype{table}\caption{#1}}
\makeatother

\begin{document}

\title{Observation of $\nu_\tau$ appearance in the CNGS beam with the OPERA experiment}

\author{
\name{N.~Agafonova}{1}, 
\name{A.~Aleksandrov}{2}, 
\name{A.~Anokhina}{3}, 
\name{S.~Aoki}{4}, 
\name{A.~Ariga}{5}, 
\name{T.~Ariga}{5,\ast}, 
\name{T.~Asada}{6}, 
\name{D.~Bender}{7}, 
\name{A.~Bertolin}{8}, 
\name{C.~Bozza}{9}, 
\name{R.~Brugnera}{8,10}, 
\name{A.~Buonaura}{2,11}, 
\name{S.~Buontempo}{2}, 
\name{B.~B\"{u}ttner}{12}, 
\name{M.~Chernyavsky}{13}, 
\name{A.~Chukanov}{14}, 
\name{L.~Consiglio}{2}, 
\name{N.~D'Ambrosio}{15}, 
\name{G.~De~Lellis}{2,11}, 
\name{M.~De~Serio}{16,17}, 
\name{P.~Del~Amo~Sanchez}{18}, 
\name{A.~Di~Crescenzo}{2,11}, 
\name{D.~Di~Ferdinando}{19}, 
\name{N.~Di~Marco}{15}, 
\name{S.~Dmitrievski}{14}, 
\name{M.~Dracos}{20}, 
\name{D.~Duchesneau}{18}, 
\name{S.~Dusini}{8}, 
\name{T.~Dzhatdoev}{3}, 
\name{J.~Ebert}{12}, 
\name{A.~Ereditato}{5}, 
\name{R.~A.~Fini}{16}, 
\name{T.~Fukuda}{21}, 
\name{G.~Galati}{16, 17}, 
\name{A.~Garfagnini}{8,10}, 
\name{G.~Giacomelli}{19,22,\thanks{Deceased}}, 
\name{C.~Goellnitz}{12}, 
\name{J.~Goldberg}{23}, 
\name{Y.~Gornushkin}{14}, 
\name{G.~Grella}{9}, 
\name{M.~Guler}{7}, 
\name{C.~Gustavino}{24}, 
\name{C.~Hagner}{12}, 
\name{T.~Hara}{4}, 
\name{T.~Hayakawa}{6}, 
\name{A.~Hollnagel}{12}, 
\name{B.~Hosseini}{2,11}, 
\name{H.~Ishida}{21}, 
\name{K.~Ishiguro}{6}, 
\name{K.~Jakovcic}{25}, 
\name{C.~Jollet}{20}, 
\name{C.~Kamiscioglu}{7,26}, 
\name{M.~Kamiscioglu}{7}, 
\name{T.~Katsuragawa}{6}, 
\name{J.~Kawada}{5}, 
\name{H.~Kawahara}{6}, 
\name{J.~H.~Kim}{27}, 
\name{S.~H.~Kim}{28}, 
\name{N.~Kitagawa}{6}, 
\name{B.~Klicek}{25}, 
\name{K.~Kodama}{29}, 
\name{M.~Komatsu}{6}, 
\name{U.~Kose}{8}, 
\name{I.~Kreslo}{5}, 
\name{A.~Lauria}{2,11}, 
\name{J.~Lenkeit}{12}, 
\name{A.~Ljubicic}{25}, 
\name{A.~Longhin}{30}, 
\name{P.~Loverre}{24,31}, 
\name{M.~Malenica}{25},
\name{A.~Malgin}{1}, 
\name{G.~Mandrioli}{19}, 
\name{T.~Matsuo}{21}, 
\name{V.~Matveev}{1}, 
\name{N.~Mauri}{19,22}, 
\name{E.~Medinaceli}{8,10}, 
\name{A.~Meregaglia}{20}, 
\name{M.~Meyer}{12}, 
\name{S.~Mikado}{32}, 
\name{M.~Miyanishi}{6}, 
\name{P.~Monacelli}{24}, 
\name{M.~C.~Montesi}{2,11}, 
\name{K.~Morishima}{6}, 
\name{M.~T.~Muciaccia}{16,17}, 
\name{N.~Naganawa}{6}, 
\name{T.~Naka}{6}, 
\name{M.~Nakamura}{6}, 
\name{T.~Nakano}{6}, 
\name{Y.~Nakatsuka}{6,\ast}, 
\name{K.~Niwa}{6}, 
\name{S.~Ogawa}{21}, 
\name{N.~Okateva}{13}, 
\name{A.~Olshevsky}{14}, 
\name{T.~Omura}{6}, 
\name{K.~Ozaki}{4}, 
\name{A.~Paoloni}{30}, 
\name{B.~D.~Park}{33}, 
\name{I.~G.~Park}{27}, 
\name{L.~Pasqualini}{19,22},
\name{A.~Pastore}{16}, 
\name{L.~Patrizii}{19}, 
\name{H.~Pessard}{18}, 
\name{C.~Pistillo}{5}, 
\name{D.~Podgrudkov}{3}, 
\name{N.~Polukhina}{13}, 
\name{M.~Pozzato}{19,22}, 
\name{F.~Pupilli}{15}, 
\name{M.~Roda}{8,10}, 
\name{T.~Roganova}{3}, 
\name{H.~Rokujo}{6}, 
\name{G.~Rosa}{24,31}, 
\name{O.~Ryazhskaya}{1}, 
\name{O.~Sato}{6}, 
\name{A.~Schembri}{15}, 
\name{I.~Shakiryanova}{1}, 
\name{T.~Shchedrina}{2}, 
\name{A.~Sheshukov}{2}, 
\name{H.~Shibuya}{21}, 
\name{T.~Shiraishi}{6}, 
\name{G.~Shoziyoev}{3}, 
\name{S.~Simone}{16,17}, 
\name{M.~Sioli}{19,22}, 
\name{C.~Sirignano}{8,10}, 
\name{G.~Sirri}{19}, 
\name{M.~Spinetti}{30}, 
\name{L.~Stanco}{8}, 
\name{N.~Starkov}{13}, 
\name{S.~M.~Stellacci}{9}, 
\name{M.~Stipcevic}{25}, 
\name{P.~Strolin}{2,11}, 
\name{S.~Takahashi}{4}, 
\name{M.~Tenti}{19}, 
\name{F.~Terranova}{30,34}, 
\name{V.~Tioukov}{2}, 
\name{S.~Tufanli}{5}, 
\name{A.~Umemoto}{6}, 
\name{P.~Vilain}{35}, 
\name{M.~Vladimirov}{13}, 
\name{L.~Votano}{30}, 
\name{J.~L.~Vuilleumier}{5}, 
\name{G.~Wilquet}{35}, 
\name{B.~Wonsak}{12}, 
\name{C.~S.~Yoon}{27}, 
\name{I.~Yaguchi}{5}, 
\name{M.~Yoshimoto}{6}, 
\name{S.~Zemskova}{14} 
\name{and A.~Zghiche}{18}
\newline
\name{(OPERA Collaboration)}{}
}

\address{
\affil{1}{INR Institute for Nuclear Research, Russian Academy of Sciences RUS-117312, Moscow, Russia}
\affil{2}{INFN Sezione di Napoli, I-80125 Napoli, Italy}
\affil{3}{SINP MSU-Skobeltsyn Institute of Nuclear Physics, Lomonosov Moscow State University, RUS-119992 Moscow, Russia}
\affil{4}{Kobe University, J-657-8501 Kobe, Japan}
\affil{5}{Albert Einstein Center for Fundamental Physics, Laboratory for High Energy Physics (LHEP), University of Bern, CH-3012 Bern, Switzerland}
\affil{6}{Nagoya University, J-464-8602 Nagoya, Japan}
\affil{7}{METU Middle East Technical University, TR-06531 Ankara, Turkey}
\affil{8}{INFN Sezione di Padova, I-35131 Padova, Italy}
\affil{9}{Dip. di Fisica dell'Uni. di Salerno and ``Gruppo Collegato" INFN, I-84084 Fisciano (SA) Italy}
\affil{10}{Dipartimento di Fisica e Astronomia dell'Universit$\grave{a}$ di Padova, I-35131 Padova, Italy}
\affil{11}{Dipartimento di Scienze Fisiche dell'Universit$\grave{a}$ Federico II di Napoli, I-80125 Napoli, Italy}
\affil{12}{Hamburg University, D-22761 Hamburg, Germany}
\affil{13}{LPI-Lebedev Physical Institute of the Russian Academy of Sciences, 119991 Moscow, Russia}
\affil{14}{JINR-Joint Institute for Nuclear Research, RUS-141980 Dubna, Russia}
\affil{15}{INFN-Laboratori Nazionali del Gran Sasso, I-67010 Assergi (L'Aquila), Italy}
\affil{16}{INFN Sezione di Bari, I-70126 Bari, Italy}
\affil{17}{Dipartimento di Fisica dell'Universit$\grave{a}$ di Bari, I-70126 Bari, Italy}
\affil{18}{LAPP, Universit\'e de Savoie, CNRS IN2P3, F-74941 Annecy-le-Vieux, France}
\affil{19}{INFN Sezione di Bologna, I-40127 Bologna, Italy}
\affil{20}{IPHC, Universit$\grave{e}$ de Strasbourg, CNRS/IN2P3, F-67037 Strasbourg, France}
\affil{21}{Toho University, J-274-8510 Funabashi, Japan}
\affil{22}{Dipartimento di Fisica e Astronomia dell'Universit$\grave{a}$ di Bologna, I-40127 Bologna, Italy}
\affil{23}{Department of Physics, Technion, IL-32000 Haifa, Israel}
\affil{24}{INFN Sezione di Roma, I-00185 Roma, Italy}
\affil{25}{IRB-Rudjer Boskovic Institute, HR-10002 Zagreb, Croatia}
\affil{26}{Ankara University, TR-06100 Ankara, Turkey}
\affil{27}{Gyeongsang National University, ROK-900 Gazwa-dong, Jinju 660-701, Korea}
\affil{28}{Center for Underground Physics, IBS, Daejeon, Korea}
\affil{29}{Aichi University of Education, J-448-8542 Kariya (Aichi-Ken), Japan}
\affil{30}{INFN-Laboratori Nazionali di Frascati dell'INFN, I-00044 Frascati (Roma), Italy}
\affil{31}{Dipartimento di Fisica dell'Universit$\grave{a}$ di Roma `La Sapienza' and INFN, I-00185 Roma, Italy}
\affil{32}{Nihon University, J-275-8576 Narashino, Japan}
\affil{33}{Samsung Changwon Hospital, SKKU, Changwon, Korea}
\affil{34}{Dipartimento di Fisica dell'Universit$\grave{a}$ di Milano-Bicocca, I-20126 Milano, Italy}
\affil{35}{IIHE, Universit\'e Libre de Bruxelles, B-1050 Brussels, Belgium}
\vspace{1mm}
\affil{*}{Corresponding authors. E-mail: tomoko.ariga@lhep.unibe.ch, nakatsuka@flab.phys.nagoya-u.ac.jp}}

\begin{abstract}

The OPERA experiment is searching for $\nu_\mu \rightarrow \nu_\tau$ oscillations in appearance mode i.e. via the direct detection of $\tau$ leptons in $\nu_\tau$ charged current interactions. The evidence of $\nu_\mu \rightarrow \nu_\tau$ appearance has been previously reported with three $\nu_\tau$ candidate events using a sub-sample of data from the 2008-2012 runs. We report here a fourth $\nu_\tau$ candidate event, with the $\tau$ decaying into a hadron, found after adding the 2012 run events without any muon in the final state to the data sample. Given the number of analysed events and the low background, $\nu_\mu \rightarrow \nu_\tau$ oscillations are established with a significance of 4.2$\sigma$. 

\end{abstract}

\subjectindex{C04, C32}

\parindent0pt

\maketitle

\section{Introduction} 

Neutrino oscillations have been studied by many experiments in disappearance mode \cite{pdg}. In 2010 the OPERA experiment, searching for $\nu_\mu \rightarrow \nu_\tau$ oscillations in appearance mode by detecting $\tau$ leptons produced in $\nu_\tau$ charged current interactions, observed a first $\nu_\tau$ candidate \cite{1tau}. Recently the Super-Kamiokande experiment reported the evidence of a $\nu_\tau$ appearance signal in atmospheric neutrino data \cite{sk2}, with a signal-to-noise ratio of about one tenth. In parallel, OPERA and T2K, both operating with accelerator-based neutrino beams, have shown the first proof of flavour transitions in appearance mode with a high signal-to-noise ratio \cite{t2k0}. While the T2K experiment has observed $\nu_\mu \rightarrow \nu_e$ oscillations \cite{t2k1}, the OPERA experiment, operating in low background conditions and with a signal-to-noise ratio of about ten, has reported the 3$\sigma$ evidence of $\nu_\mu \rightarrow \nu_\tau$ appearance \cite{3tau} using a subsample of the data collected in the 2008-2012 runs. In this paper, we describe an additional $\nu_\tau$ candidate event, with the $\tau$ decaying into a hadron, found in the analysis of an extended data set, including also the events of the 2012 run without any muon in the final state. We finally report the achieved significance in the observation of $\nu_\mu \rightarrow \nu_\tau$ oscillations.

\section{The OPERA detector and the data sample}
In order to meet the experimental requirements of a large target mass and a micrometric spatial accuracy to detect short-lived $\tau$ lepton decays, a target made of lead plates interspaced with emulsion films acting as tracking devices is used. The OPERA detector \cite{detector} is composed of two identical parts called Super Modules, each consisting of a target section followed by a spectrometer. The target has an average mass of about 1.2 kt and a modular structure with an average number of 141431 target units, called bricks. A brick consists of 57 emulsion films \cite{film}, interleaved with 1 mm thick lead plates. Bricks are arranged in walls alternated with scintillator strip planes (Target Tracker or TT). Magnetic spectrometers, consisting of iron core magnets instrumented with Resistive Plate Chambers (RPC) and drift tubes (high-Precision Tracker or PT), are used for the measurement of the muon charge \cite{charge} and momentum. In order to reduce the emulsion scanning load, removable pairs of emulsion films called Changeable Sheets \cite{cs} are used as interface trackers between the TT and the bricks. 

The OPERA detector was exposed to the CERN Neutrinos to Gran Sasso (CNGS) beam \cite{cngs1,cngs2} from 2008 to 2012. A sample of 19505 contained neutrino interactions corresponding to 17.97$\times$10$^{19}$ protons on target (pot) have been registered by the detector. A three-dimensional track is tagged as a muon if the product of its length and the density along its path is larger than 660 g/cm$^2$ \cite{eledet}. An event is classified as 1$\mu$ if either it contains at least one track tagged as a muon or the total number of fired TT or RPC planes is larger than 19. The complementary sample is defined as 0$\mu$. 0$\mu$ events contain the signals of $\tau \rightarrow 1h$, $\tau \rightarrow 3h$ and $\tau \rightarrow e$ decay channels. 1$\mu$ events contain the $\tau \rightarrow \mu$ decay channel. Most of signal events occur at low muon momentum thus a muon momentum cut at 15 GeV/c was introduced to accelerate the finding of signal for $\nu_\mu \rightarrow \nu_\tau$ oscillations. 

The data set used in the present analysis consists of the 0$\mu$ events and the 1$\mu$ events with a muon momentum smaller than 15 GeV/c that were collected during the 5 years of run. With respect to our most recent study \cite{3tau}, it now includes the last missing sample, the 0$\mu$ events of the 2012 run. For runs 2008-2009, all events have been searched for in the two most probable bricks while so far only in the first most probable brick for runs 2010-2012. The numbers of analysed events are summarised in Table \ref{tb:analysed}. 

The selection criteria of $\nu_\tau$ interactions and the evaluation of efficiencies and backgrounds are described in detail in \cite{2tau}. The observation of three $\nu_\tau$ candidate events has been reported in previous articles \cite{1tau,2tau,3tau}. A new $\nu_\tau$ candidate event was found in this new sample and is reported here. 

\renewcommand{\arraystretch}{1.0}
\begin{table}[hbtb]
\begin{center}
\begin{tabular}{|l|c|c|c|c|c|c|}
\hline
                                    & 2008 & 2009 & 2010 & 2011 & 2012 & Total \\
\hline
pot (10$^{19}$)                     & 1.74 & 3.53 & 4.09 & 4.75 & 3.86 & 17.97 \\
\hline
0$\mu$ events                       &  148 &  250 &  209 &  223 &  149 &   979 \\
\hline
1$\mu$ events ($p_\mu$ $<$ 15 GeV/c)&  534 & 1019 &  814 &  749 &  590 &  3706 \\
\hline
Total of events                     &  682 & 1269 & 1023 &  972 &  739 &  4685 \\
\hline
\end{tabular}
\caption{Number of events used in this analysis \protect \footnotemark. 
}
\label{tb:analysed}
\end{center}
\end{table}
\renewcommand{\arraystretch}{1}
\footnotetext{\scriptsize 
An additional sample of 835 1$\mu$ events with a muon momentum larger than 15 GeV/c from runs 2008-2009 has also been analysed for the purpose of probing our understanding of the beam spectrum and the detector response. They are not included in the data set used for this paper. The grand total number of fully analysed events is therefore 5520.
}
\normalsize

\section{Description of the new $\nu_\tau$ candidate event}

\begin{figure}[htbp]
\begin{center}
\includegraphics[width=\linewidth]{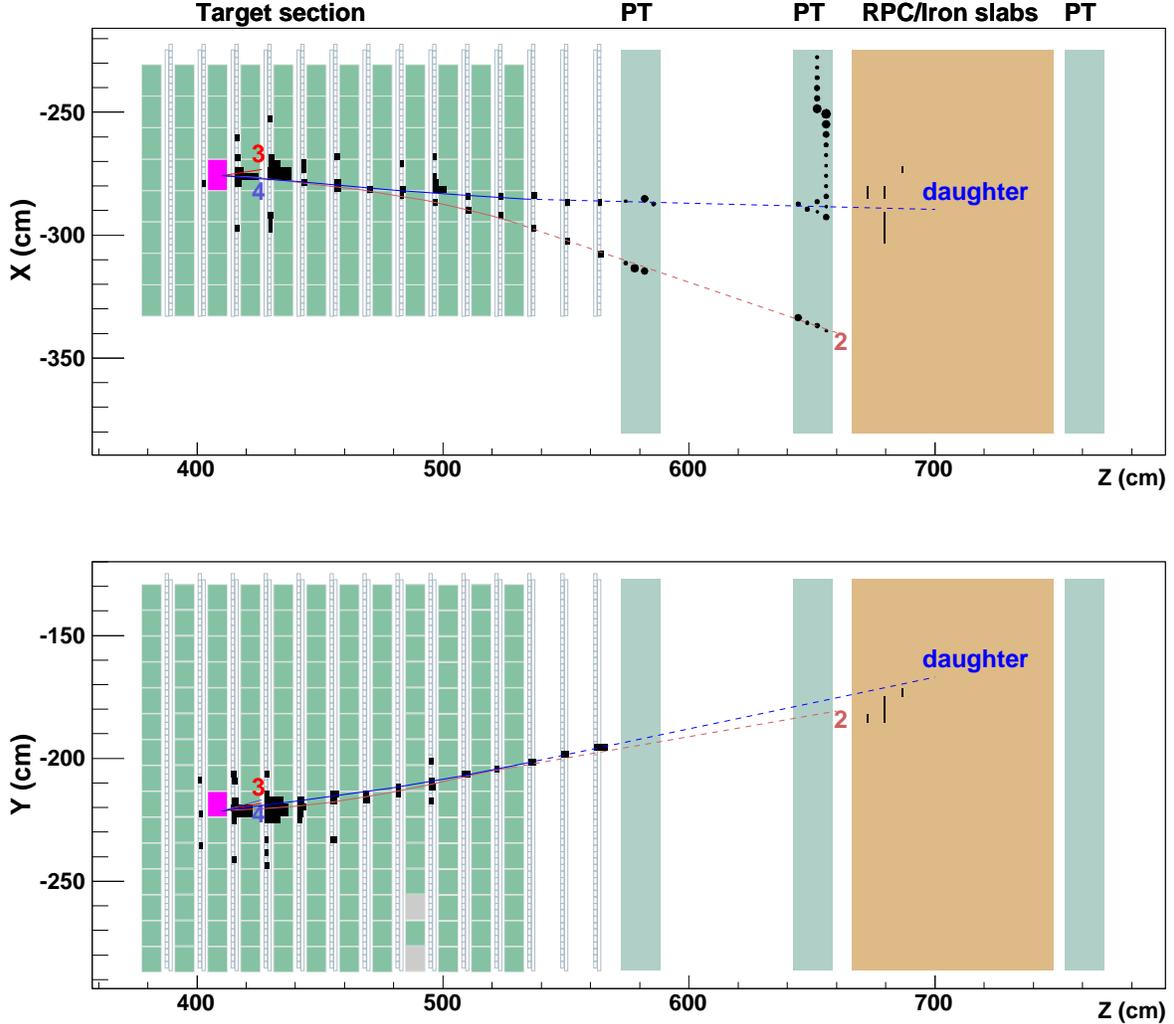}
\end{center}
\caption{
Display of the $\nu_\tau$ candidate event as seen by the electronic detectors in the $x$-$z$ projection (top) and $y$-$z$ projection (bottom). Neutrinos are coming from the left side. The brick containing the neutrino interaction is highlighted in magenta. Solid lines show the position of tracks measured in the primary and downstream bricks. Dashed lines show the linear extrapolation of the tracks using positions and slopes at the last measured point in the bricks. Labelled tracks are discussed in the text. The pattern of hits starting from the daughter track visible in the $x$-$z$ projection in the PT is consistent with a backscattered proton or pion from an interaction of the daughter track. 
} \label{fig:ttdisplay}
\end{figure}

This neutrino interaction occurred on 9~September 2012 in the second Super Module, ten brick walls upstream of the spectrometer. The event is classified as 0$\mu$. Fig.~\ref{fig:ttdisplay} shows a display of the event as seen by the electronic detectors. 

\begin{figure}[htbp]
\begin{center}
\begin{minipage}{0.49\textwidth}
\begin{center}
\includegraphics[width=\textwidth]{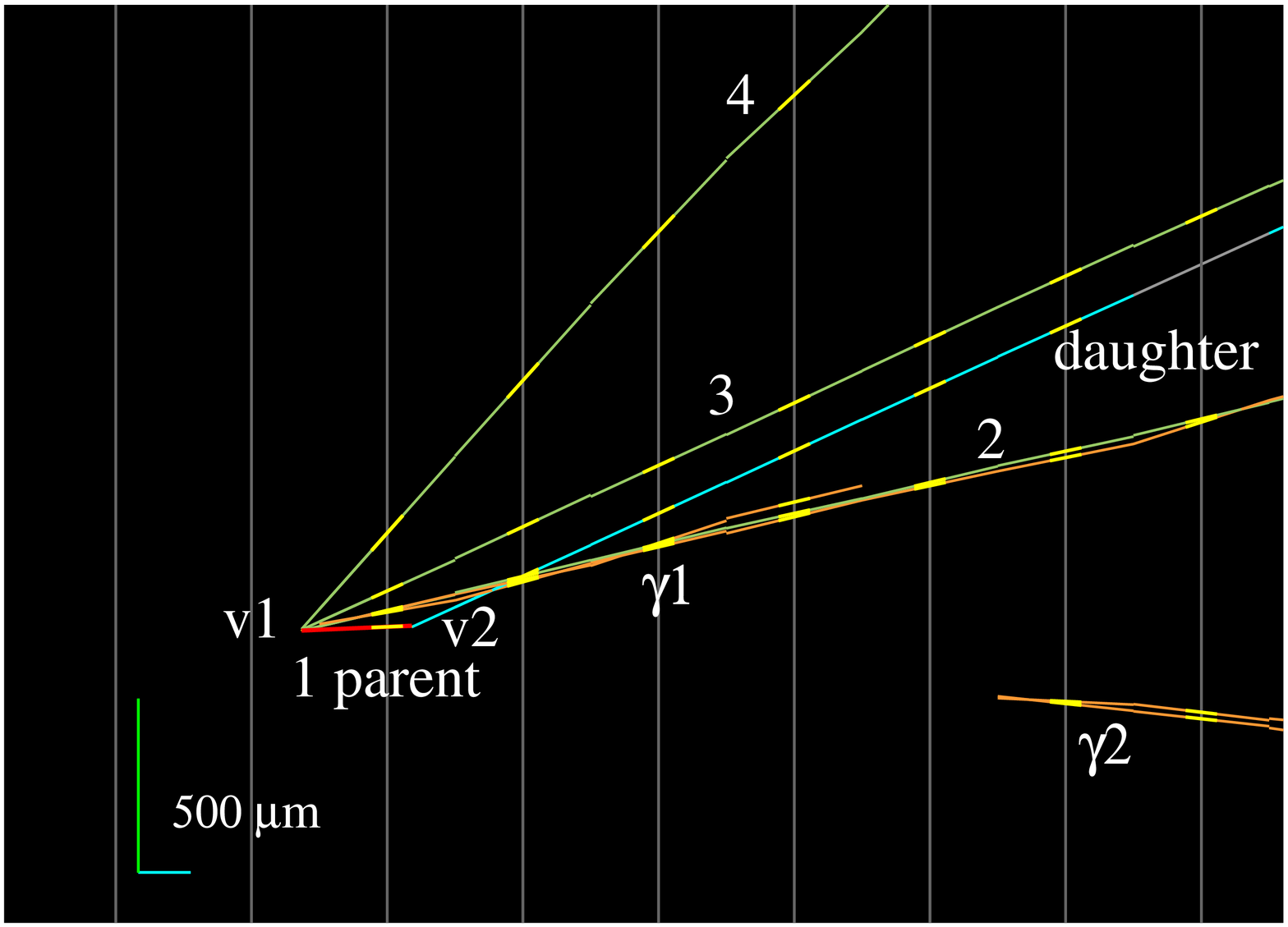}
\end{center}
\end{minipage}
\begin{minipage}{0.49\textwidth}
\begin{center}
\includegraphics[width=\textwidth]{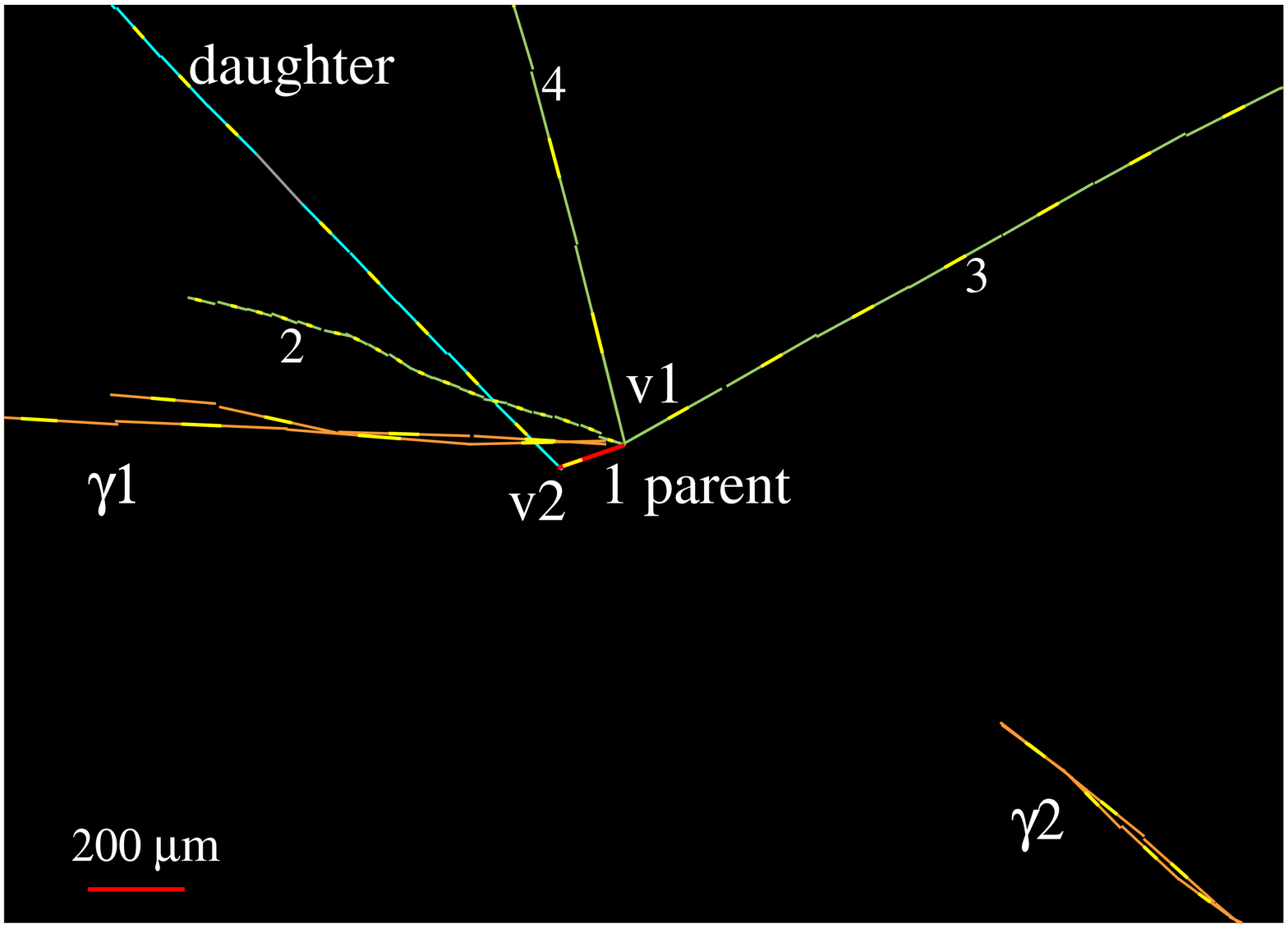}
\end{center}
\end{minipage}
\caption{Event display of the fourth $\nu_\tau$ candidate event in the $y$-$z$ projection longitudinal to the neutrino direction (left) and in the view transverse to the neutrino direction (right). The primary and decay vertices are indicated as ``v1" and ``v2" respectively. Labelled tracks are discussed in the text.
} \label{fig:display}
\end{center}
\end{figure}

A converging pattern of 10 tracks was found in the Changeable Sheets and the primary vertex was located, 18 plates from the downstream face. The primary vertex consists of four tracks of which one exhibits a kink topology (decay vertex). No nuclear fragment associated to the decay vertex was found by a dedicated scanning procedure with an extended angular acceptance (up to $\tan \theta$ = 3.5, $\theta$ being the angle of the track with respect to the $z$ axis) \cite{anis,toho}. Two electromagnetic showers initiated by conversions of $\gamma$-rays have been detected, both pointing to the primary vertex as described below. Fig.~\ref{fig:display} shows a display of this event as reconstructed in the brick. Momenta of the reconstructed tracks are determined by Multiple Coulomb Scattering (MCS) in the bricks \cite{mom}. 

\begin{itemize}
\item Track 1 is the parent track of a kink topology with an angle of (137 $\pm$ 4) mrad. The flight length is (1090 $\pm$ 30) $\mu$m. The longitudinal coordinate of the decay vertex with respect to the downstream face of the lead plate containing the primary vertex ($z_{dec}$) is (406 $\pm$ 30) $\mu$m. The kink angle and $z_{dec}$ satisfy the topological selection for $\nu_\tau$ interaction search, which are $\theta_{kink}$ $>$ 20 mrad and $z_{dec}$ $<$ 2600 $\mu$m. This track is the $\tau$ candidate. 

\item Track 2 has a momentum of (1.9$^{+0.3}_{-0.2}$) GeV/c and was followed in the downstream bricks till the end of the target. It is exiting the target and entering the spectrometer leaving hits in the PT but not in the RPC as can be seen from Fig.~\ref{fig:ttdisplay}. 

\item Track 3 has a momentum of (1.1$^{+0.2}_{-0.1}$) GeV/c and has an interaction just before entering the 2$^{nd}$ downstream brick, producing two charged tracks in the brick. 

\item Track 4 is a heavily ionising particle having $p\beta$ = (0.4 $\pm$ 0.1) GeV/c. It stops between the 1$^{st}$ and 2$^{nd}$ downstream bricks. From its range, (94 $\pm$ 1) g/cm$^{2}$, the particle is identified as a proton with a momentum of (0.7 $\pm$ 0.1) GeV/c. 

\item The kink daughter track has a momentum of (6.0$^{+2.2}_{-1.2}$) GeV/c. Its impact parameter with respect to the primary vertex is (146 $\pm$ 5) $\mu$m. It was followed in the downstream bricks till the end of the target. It is exiting the target, stopping in the spectrometer after leaving a signal in three RPC planes (Fig.~\ref{fig:ttdisplay}).

\end{itemize}

None of the charged particles at both vertices is identified as an electron due to the absence of electromagnetic showers. All tracks were followed down in the downstream bricks and a total of 20 bricks were analysed. 

\begin{figure}[htbp]
\begin{center}
\includegraphics[width=0.5\textwidth]{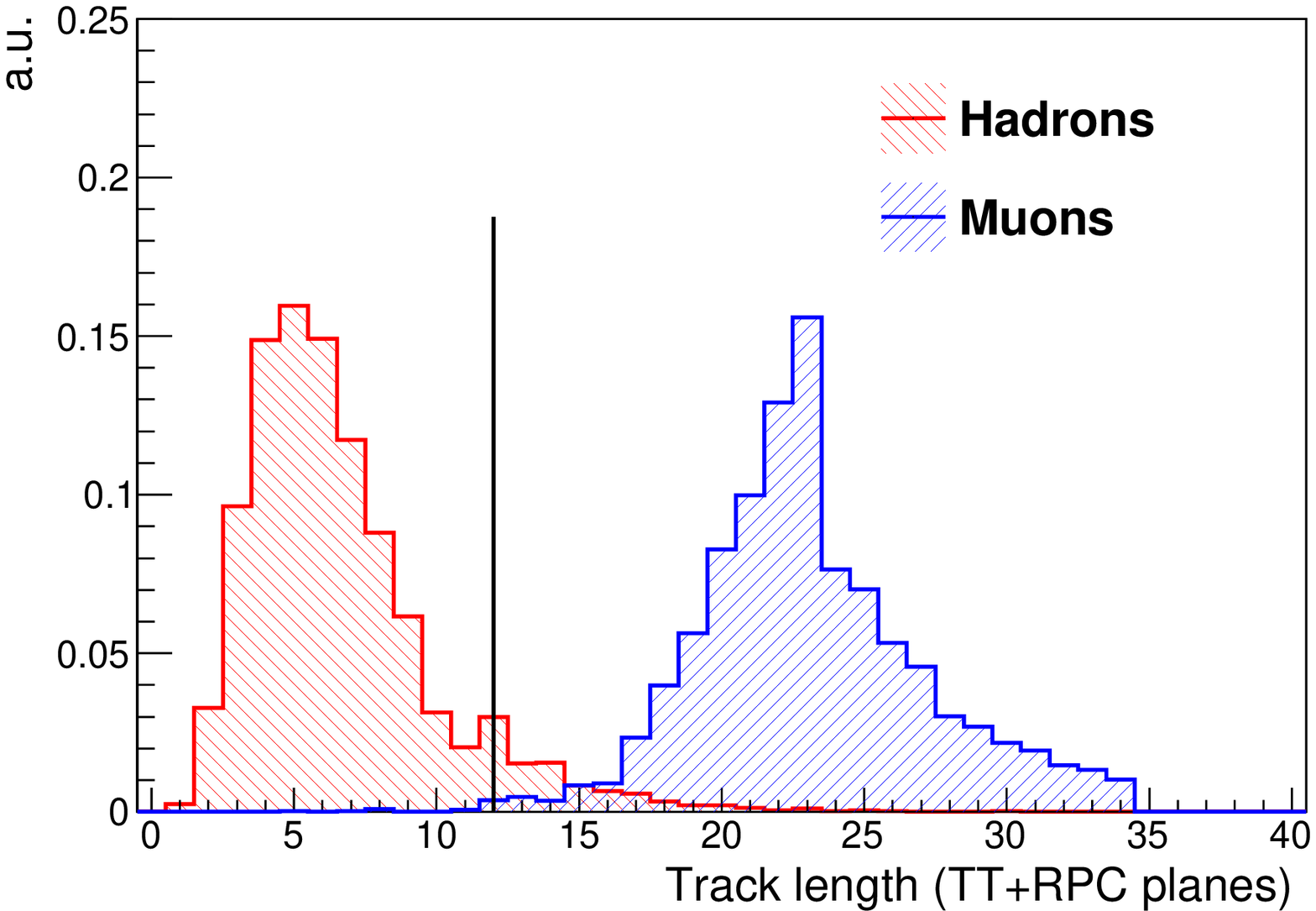}
\caption{Monte Carlo distribution of track lengths for hadrons and muons originated in the same target position as the observed event and with the same momentum as measured for track 2. The black vertical line shows the value for track 2. } \label{fig:track_length}
\end{center}
\end{figure}

The momentum of track 2 was measured at each downstream brick over 10 bricks and the values were combined to estimate the momentum at the primary vertex. Track 2 is not classified as a muon since the total material crossed is 604 g/cm$^2$, which is below the lower cut for $\mu$ identification set at 660 g/cm$^2$. To separate muons from hadrons, momentum-range correlations are characterised by a discriminating variable $D_{TFD} = \frac{L}{R(p)}$ $\frac{\langle \rho \rangle}{\rho}$ where $L$ is the track length, $R(p)$ is the range in lead of a muon with momentum $p$, $\langle \rho \rangle$ is the average density along the path and $\rho$ is the lead density \cite{2tau}. A track is classified as a muon if $D_{TFD}$ is above 0.8 while for track 2 we have $D_{TFD} = $0.40$^{+0.04}_{-0.05}$. Thus the muon hypothesis for track 2 is rejected. A further test of the hadron/muon hypothesis was performed using the track length in the TT and the RPC \cite{alessandro}. Track 2 crosses 12 planes. Fig.~\ref{fig:track_length} shows the Monte Carlo distribution of track lengths for hadrons and muons with momentum (1.9$^{+0.3}_{-0.2}$) GeV/c originated in the same target position as the observed event. The probability for a muon to cross less than 13 planes is 0.4\% while the probability for a pion to cross more than 11 planes is 9.2\%. This result confirms the rejection of the muon hypothesis. 

Track 3 is identified as a hadron due to the presence of an interaction. The kink daughter track is identified as a hadron based on $D_{TFD}$ = 0.18 $\pm$ 0.04. 

The conversion point of $\gamma$-ray 1 is in the lead plate containing the primary vertex, 376 $\mu$m upstream of the decay vertex. The energy of $\gamma$-rays are measured by MCS of the $e^+$ and $e^-$ pair \cite{mom} taking their energy loss into account. The $\gamma$-ray 1 has an energy of (0.7$^{+0.2}_{-0.1}$) GeV and its impact parameter to the primary vertex is (2$^{+8}_{-2}$) $\mu$m. 

The conversion point of $\gamma$-ray 2 is 5 lead plates downstream of the primary vertex. $\gamma$-ray 2 has an energy of (4.0$^{+4.7}_{-1.4}$) GeV. Its impact parameter is (33$^{+43}_{-33}$) $\mu$m with respect to the primary vertex while it is (267 $\pm$ 36) $\mu$m with respect to the decay vertex. The probability that it originates from the decay vertex is less than 10$^{-3}$. 
Both $\gamma$-rays are not from the $\tau$ decay and the candidate is $\tau \rightarrow 1h$ decay. 
The invariant mass of $\gamma$-ray 1 and $\gamma$-ray 2 is (0.59$^{+0.20}_{-0.15}$) GeV/c$^2$, indicating that they are not the decay products of the same $\pi^0$. 

The scalar sum of the momenta of all particles measured in the emulsion films ($p_{sum}$) is (14.4$^{+3.9}_{-2.7}$) GeV/c. 

\renewcommand{\arraystretch}{1.2}
\begin{figure}[htbp]
\begin{center}
\begin{minipage}{0.37\textwidth}
\vspace{5mm}
\includegraphics[width=\textwidth]{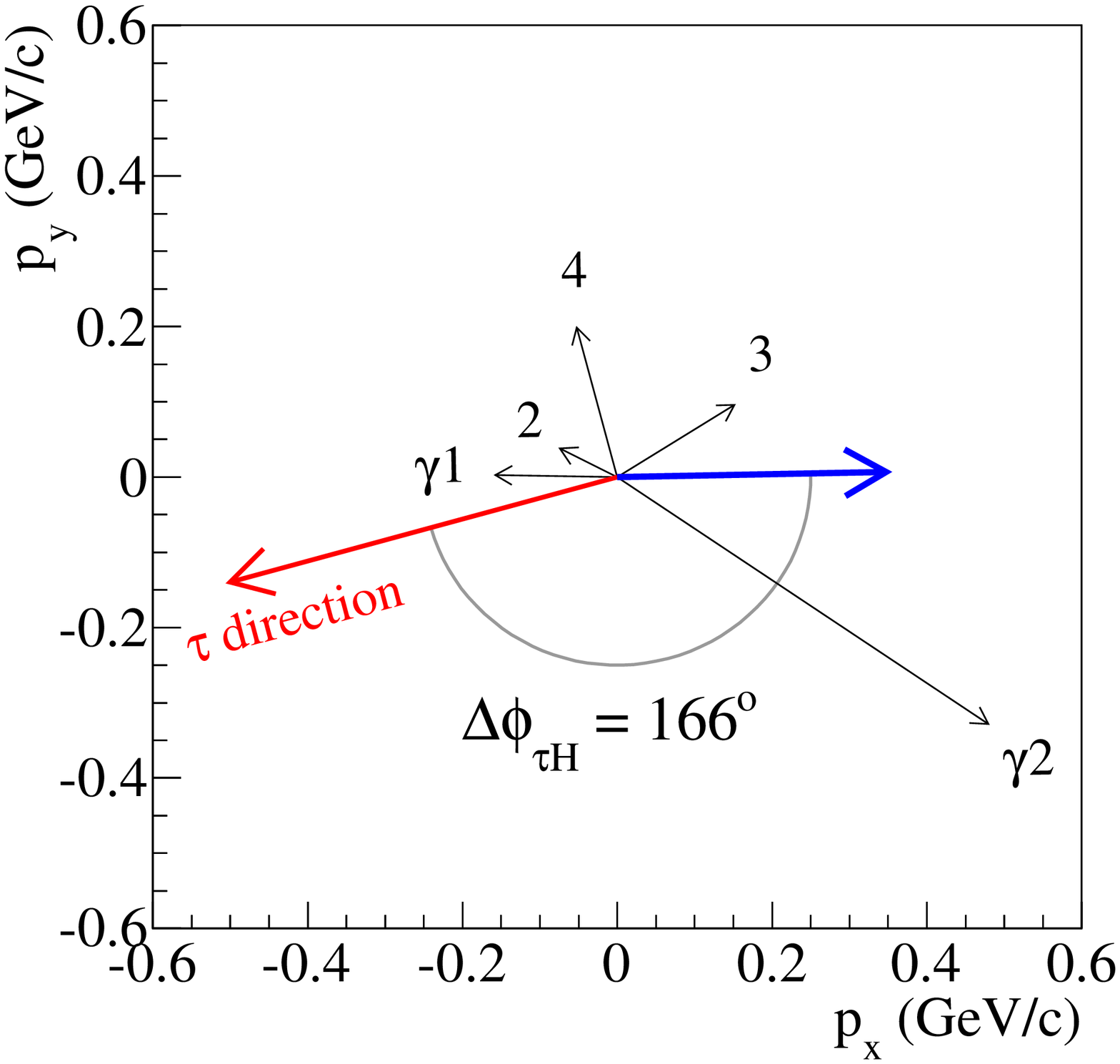}
\caption{The $\tau$ direction (red arrow) and the other primary particles (black arrows) in the plane transverse to the beam. 
The blue arrow shows the vectorial sum of the primary particles except the parent. 
} \label{fig:phi}
\end{minipage}
\hspace{5mm}
\begin{minipage}{0.57\textwidth}
\makeatletter
\def\@captype{table}
\makeatother
\begin{tabular}{ccc}
\hline
Variable & Selection & Measured value\\
\hline
$\theta_{kink}$ (mrad)         & $>$ 20  & $137 \pm 4$            \\
$z_{dec}$ ($\mu$m)             & $<$ 2600& $406 \pm 30$           \\
$p_{2ry}$ (GeV/c)              & $>$ 2   & $6.0^{+2.2}_{-1.2}$    \\
$p_T^{2ry}$ (GeV/c)            & $>$ 0.6 (0.3$^*$) & $0.82^{+0.30}_{-0.16}$ \\
$p_T^{miss}$ (GeV/c)           & $<$ 1   & $0.55^{+0.30}_{-0.20}$ \\
$\Delta\phi_{\tau H}$ (degrees)& $>$ 90  & $166^{+2}_{-31}$       \\
\hline
\end{tabular}
\vspace{5mm}
\caption{Selection criteria for $\nu_\tau$ interaction search in the $\tau \rightarrow 1h$ decay channel and the values measured for the fourth $\nu_\tau$ candidate event. Cut marked with $^*$ is applied if there is at least one $\gamma$-ray originating from the decay vertex.}
\label{tb:variables}
\end{minipage}
\end{center}
\end{figure}
\renewcommand{\arraystretch}{1}

\begin{figure}[htbp]
\begin{center}
\includegraphics[width=1.0\linewidth]{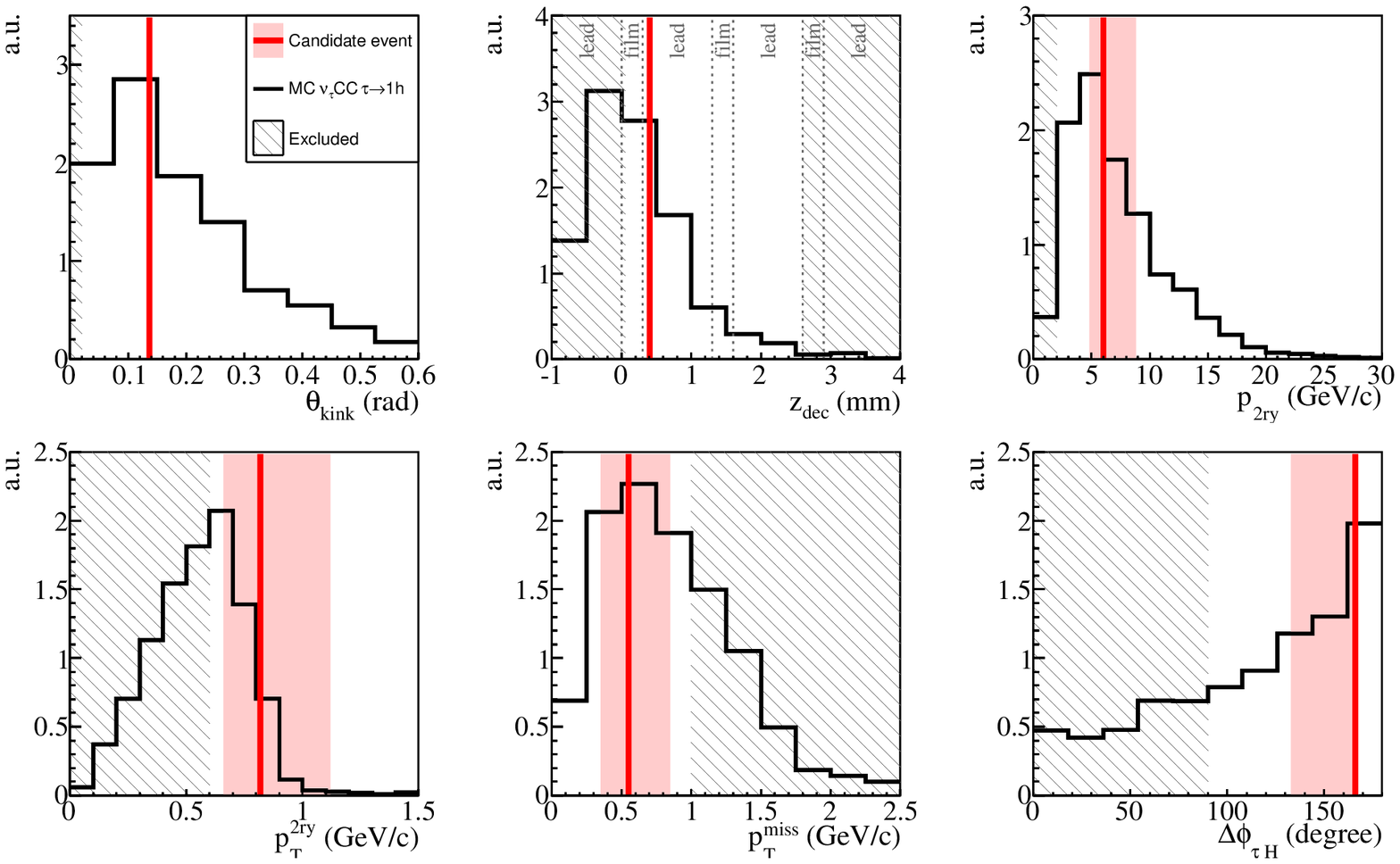}
\end{center}
\caption{Monte Carlo distributions of the kinematical variables for $\nu_{\tau}$ events passing all the cuts in the $\tau \rightarrow 1h$ decay channel. 
Red lines show the measured values for the candidate event reported here and red band their error. Grey areas show the region excluded by the selection criteria. 
} \label{fig:allfig}
\end{figure}

The momentum of the daughter track ($p_{2ry}$) is (6.0$^{+2.2}_{-1.2}$) GeV/c, well above the cut value of 2 GeV/c \cite{2tau}. The transverse momentum ($p_T^{2ry}$) at the decay vertex is (0.82$^{+0.30}_{-0.16}$) GeV/c, which is above the lower cut of 0.6 GeV/c. The missing transverse momentum at the primary vertex ($p_T^{miss}$) is (0.55$^{+0.30}_{-0.20}$) GeV/c, thus below the maximum allowed value which is set at 1 GeV/c. As shown in Fig.~\ref{fig:phi}, the angle between the $\tau$ candidate direction and the sum of the transverse momenta of the other primary particles ($\Delta\phi_{\tau H}$) is (166$^{+2}_{-31}$) degrees, largely above the lower cut at 90 degrees. The values of the kinematical variables for this event are summarised in Table \ref{tb:variables}. The Monte Carlo distributions of the variables and the measured values are shown in Fig.~\ref{fig:allfig}. The measured values are well within the expected signal region. 

\section{Results}

The estimated signal and background for the observation of $\nu_\tau$ candidates in the data sample analysed in this paper are obtained as described in \cite{2tau} and summarised in Table \ref{tb:numbers}. The systematic uncertainties are estimated to be 20\% on the signal, 20\% on the charm background, 30\% on the hadronic background and 50\% on the large-angle muon scattering background, 
i.e. muon scatterings mimicking $\tau \rightarrow \mu$ decays. 
The expected signal consists of (2.11 $\pm$ 0.42) $\nu_{\tau}$ events in all decay channels, using $\Delta m^{2}_{23}$ = 2.32 $\times$ 10$^{-3}$ eV$^{2}$ and $\sin ^{2}2\theta_{23}$ = 1. The total expected background for the sample is (0.233 $\pm$ 0.041) events. 

\begin{figure}[htbp]
\begin{center}
\includegraphics[width=0.55\textwidth]{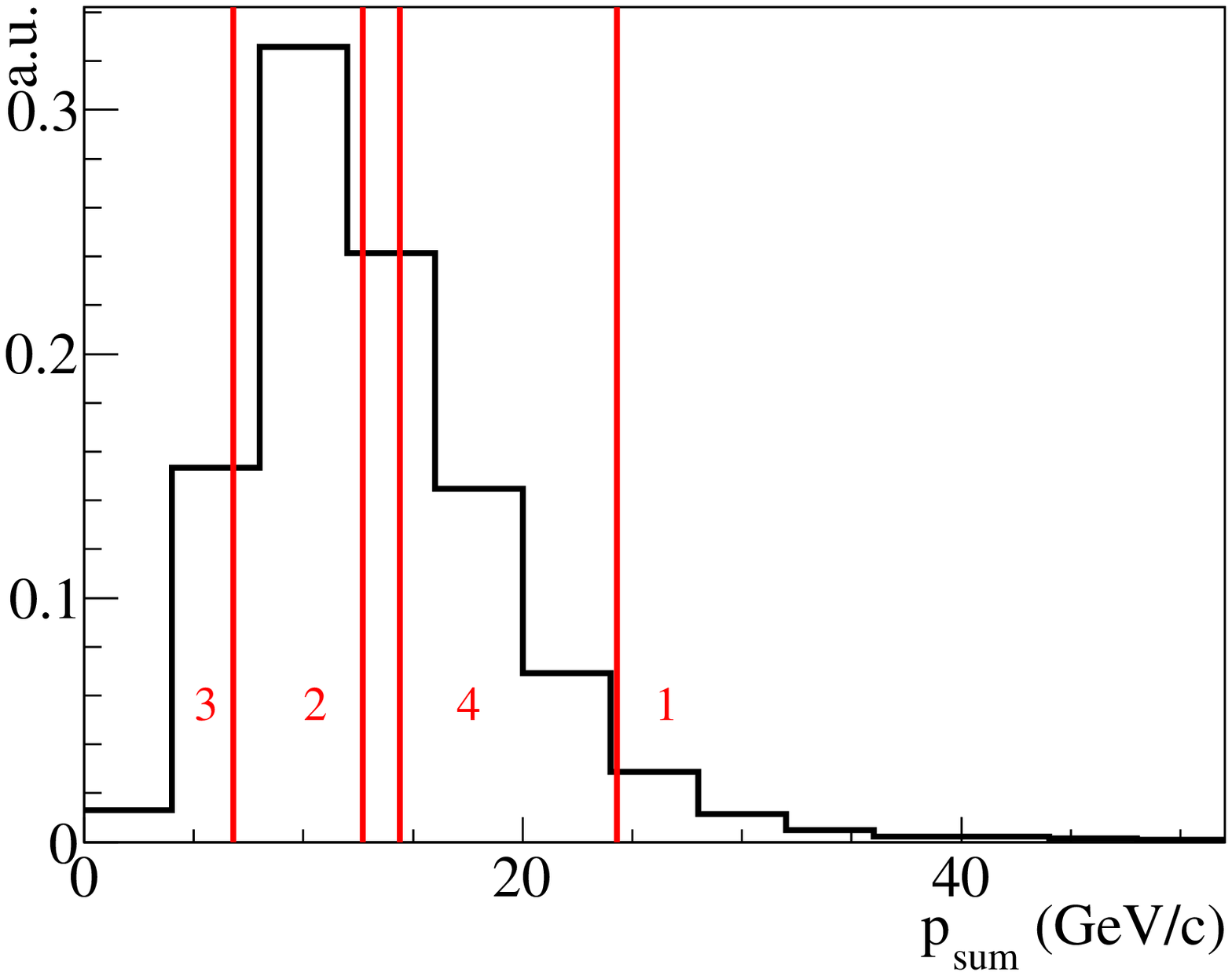}
\caption{Monte Carlo distribution of the scalar sum of the momenta of all particles measured in the emulsion films, $p_{sum}$, for $\nu_{\tau}$ events passing all the cuts in all decay channels. 
Red lines show the measured values for the four $\nu_\tau$ candidate events: ``1" and ``4" in the $\tau \rightarrow 1h$ decay channel, ``2" in $\tau \rightarrow 3h$ and ``3" in $\tau \rightarrow \mu$ \cite{1tau,2tau,3tau}. 
} \label{fig:psum_4ev}
\end{center}
\end{figure}

Four $\nu_\tau$ candidate events have been detected in the analysed samples: two in the $\tau \rightarrow 1h$ decay channel, one in $\tau \rightarrow 3h$ and one in $\tau \rightarrow \mu$ \cite{1tau,2tau,3tau}. 
The values of $p_{sum}$ measured for the four events are compatible with the corresponding signal Monte Carlo distribution as shown in Fig.~\ref{fig:psum_4ev}. 

\renewcommand{\arraystretch}{1.0}
\begin{table}[hbtb]
\footnotesize
\begin{center}
\begin{tabular}{|c|c|c|c|c|c|c|}
\hline
\multicolumn{1}{|c|}{Decay} & \multicolumn{1}{c|}{Expected} & \multicolumn{1}{c}{} & \multicolumn{4}{|c|}{Expected background} \\
\cline{4-7}
channel                & signal          & Observed & Total             & Charm             & Hadronic          & Large-angle       \\
                       &                 &          &                   & decays            & re-interactions   & muon scattering   \\
\hline
$\tau \rightarrow 1h$  & 0.41 $\pm$ 0.08 & 2        & 0.033 $\pm$ 0.006 & 0.015 $\pm$ 0.003 & 0.018 $\pm$ 0.005 & $/$               \\
$\tau \rightarrow 3h$  & 0.57 $\pm$ 0.11 & 1        & 0.155 $\pm$ 0.030 & 0.152 $\pm$ 0.030 & 0.002 $\pm$ 0.001 & $/$               \\
$\tau \rightarrow \mu$ & 0.52 $\pm$ 0.10 & 1        & 0.018 $\pm$ 0.007 & 0.003 $\pm$ 0.001 & $/$               & 0.014 $\pm$ 0.007 \\
$\tau \rightarrow e$   & 0.62 $\pm$ 0.12 & 0        & 0.027 $\pm$ 0.005 & 0.027 $\pm$ 0.005 & $/$               & $/$               \\
\hline
Total                  & 2.11 $\pm$ 0.42 & 4        & 0.233 $\pm$ 0.041 & 0.198 $\pm$ 0.040 & 0.021 $\pm$ 0.006 & 0.014 $\pm$ 0.007 \\
\hline
\end{tabular}
\caption{Estimated signal and background for the analysed sample and the number of observed events. 
}
\label{tb:numbers}
\end{center}
\normalsize
\end{table}
\renewcommand{\arraystretch}{1}

The significance of the observation of the four $\nu_\tau$ candidate events is estimated by considering the confidence for excluding the null hypothesis. Individual $p$-values of the $\tau$ decay channels are combined independently according to the Fisher's rule into an estimator $p^{*} = p_{h}p_{3h}p_{\mu}p_{e}$ \cite{sato,cdf}. In order to take into account the systematic uncertainties of the backgrounds, 100 sets of randomised backgrounds are generated. A mean $p$-value of 1.24 $\times$ 10$^{-5}$ is obtained by Monte Carlo calculation of the tail probability corresponding to the observed value of $p^{*}$. The absence of signal can be excluded with a significance of 4.2$\sigma$.

Alternatively a hypothesis test employing a likelihood-based approach \cite{likelihood} was performed. The likelihood function is $\mathcal{L} (\mu) = \prod^{4}_{i=1}e^{-(\mu s_i + b_i)} (\mu s_i + b_i)^{n_i} / n_{i}!$. The index $i$ runs over decay channels, the parameter $\mu$ determines the strength of the signal process ($\mu$ = 0 corresponds to the background-only hypothesis), $s_i$ and $b_i$ are the numbers of expected signal and background events, $n_i$ the number of observed events. The systematic uncertainties of the backgrounds were taken into account in a similar way as above. A $p$-value of 1.03 $\times$ 10$^{-5}$ corresponding to a significance of 4.2$\sigma$ for the exclusion of the null hypothesis is obtained. 

Given the 4 observed events and the expected background of (0.233 $\pm$ 0.041) events, the confidence interval of $\Delta m^2_{23}$ is estimated with the Feldman-Cousins method \cite{FC}, assuming maximal mixing. The systematic uncertainties of signal and background are taken into account to marginalise the likelihood function used for the ordering principle. The 90\% confidence interval of $\Delta m^2_{23}$ is [1.8, 5.0] $\times$ 10$^{-3}$ eV$^2$. An alternative analysis employing a Bayesian approach \cite{pdg} with a flat prior on $\Delta m^2_{23}$ was performed. The credible interval of $\Delta m^2_{23}$ is [1.9, 5.0] $\times$ 10$^{-3}$ eV$^2$. More precise measurements by other experiments \cite{k2k2,sk3,t2k2,minos2,icecube,antares} are within these intervals. 

\section{Conclusions and prospects}

A new $\nu_\tau$ candidate event, with the $\tau$ decaying into a hadron, found after adding an extended data sample is reported. Given the analysed sample, $\nu_\mu \rightarrow \nu_\tau$ oscillations are established at the 4.2$\sigma$ level. The search for events, not found in the most probable bricks in the 2010-2012 runs, is being extended to the second most probable bricks for future results. 

\ack

We thank CERN for the successful operation of the CNGS facility and INFN for the continuous support given to the experiment through its LNGS laboratory. We acknowledge funding from our national agencies: Fonds de la Recherche Scientifique-FNRS and Institut InterUniversitaire des Sciences Nucl\'eaires for Belgium; MoSES for Croatia; CNRS and IN2P3 for France; BMBF for Germany; INFN for Italy; JSPS, MEXT, QFPU-Global COE programme of Nagoya University, and Promotion and Mutual Aid Corporation for Private Schools of Japan for Japan; SNF, the University of Bern and ETH Zurich for Switzerland; the Russian Foundation for Basic Research (Grant No. 12-02-12142 ofim), the Programs of the Presidium of the Russian Academy of Sciences (Neutrino physics and Experimental and theoretical researches of fundamental interactions), and the Ministry of Education and Science of the Russian Federation for Russia, the National Research Foundation of Korea Grant (NRF-2013R1A1A2061654) for Korea and TUBITAK, the Scientific and Technological Research Council of Turkey for Turkey. We thank the IN2P3 Computing Centre (CC-IN2P3) for providing computing resources.

\end{document}